\pdfoutput=1

\documentclass[11pt]{article}

\usepackage[preprint]{acl}

\usepackage{times}
\usepackage{latexsym}

\usepackage[T1]{fontenc}

\usepackage[utf8]{inputenc}

\usepackage{microtype}
\usepackage{url}
\usepackage{inconsolata}

\usepackage{enumitem}

\usepackage{algorithm}
\usepackage{algorithmic}
\usepackage{graphicx}
\usepackage{booktabs}
\usepackage{algorithm}
\usepackage{algorithmic}
\usepackage{amsfonts,amssymb}
\usepackage{amsmath,bm}
\usepackage{subcaption}
\usepackage{diagbox}
\usepackage{url}
\usepackage{multirow} 
\usepackage{multicol} 
\usepackage{xcolor}
\usepackage{colortbl}
\usepackage{mathrsfs}
\usepackage{tabularx}
\usepackage{makecell}
\usepackage{amsfonts}
\usepackage{bbm}
\usepackage{arydshln}
\usepackage{booktabs}
\usepackage{bbding}
\usepackage{pifont}

\title{TwiUSD: A Benchmark Dataset and Structure-Aware LLM Framework for User Stance Detection
}

\author{Fuqiang Niu$^{1}$\thanks{Equal contribution.}, Zini Chen$^{1}$\footnotemark[1], Zhiyu Xie$^{1}$\footnotemark[1], Hu Huang$^{2}$, Genan Dai$^{1}$\thanks{Corresponding author.} \and  Bowen Zhang$^{1}$\footnotemark[2] \\
        $^{1}$The College of Big Data and Internet, Shenzhen Technology University, China \\
        $^{2}$School of Cyberspace Security, University of Science and Technology of China, China \\
        nfq729@gmail.com \ \ \ \ \{202100203091, 202200201110\}@stumail.sztu.edu.cn \\
        huanghu@mail.ustc.edu.cn  \ \ \ \  daigenan@sztu.edu.cn  \ \ \ \  zhang\_bo\_wen@foxmail.com\\
        }
\begin{document}
\maketitle
\begin{abstract}
User-level stance detection (UserSD) remains challenging due to the lack of high-quality benchmarks that jointly capture linguistic and social structure. In this paper, we introduce TwiUSD, the first large-scale, manually annotated UserSD benchmark with explicit followee relationships, containing 16,211 users and 47,757 tweets. TwiUSD enables rigorous evaluation of stance models by integrating tweet content and social links, with superior scale and annotation quality. Building on this resource, we propose MRFG: a structure-aware framework that uses LLM-based relevance filtering and feature routing to address noise and context heterogeneity. MRFG employs multi-scale filtering and adaptively routes features through graph neural networks or multi-layer perceptrons based on topological informativeness. Experiments show MRFG consistently outperforms strong baselines—including PLMs, graph-based models, and LLM prompting—in both in-target and cross-target evaluation. 
\end{abstract}

\section{Introduction}
Stance detection, which aims to automatically identify attitudes in text, is increasingly crucial as social media becomes the main arena for public discourse. With billions of users on platforms like Twitter, social media provides unprecedented insight into public sentiment on social, political, and cultural issues~\cite{aldayel2021stance,niu2024challenge}. Analyses of election events and policy debates highlight the value of social media data for understanding and tracking societal attitudes. Accordingly, accurate stance detection has become a cornerstone of opinion mining, with broad applications in marketing, policy, and beyond~\cite{kuccuk2020stance,LiSSNIC21}.

Research on stance detection has focused on two main levels: \textit{tweet-level}, where individual posts are classified with respect to a target, and \textit{user-level} (UserSD), where the objective is to infer a user’s overall stance by integrating signals from their textual and behavioral activity~\cite{zhang2024survey}. 
UserSD is particularly valuable for understanding opinion dynamics at both individual and community scales, as it leverages richer features such as tweets, profiles, and social connections to enable more context-aware representations.

Despite its importance, progress in UserSD is severely hampered by the lack of high-quality, large-scale benchmarks.
Existing resources suffer from two critical issues: (1) heavy reliance on distant supervision or heuristic labeling (e.g., retweet patterns or hashtag overlaps), which introduces noisy and unreliable annotations~\cite{DBLP:conf/icwsm/DarwishSAN20, ZHU2020102031, samih-darwish-2021-topical, 10.1007/978-3-031-45275-8_7,DBLP:conf/icwsm/ZhangZP024}; and (2) a tweet-centric design that aggregates tweet-level annotations without modeling the essential social structure, such as follower or mention relationships~\cite{10.1007/978-3-031-17091-1_43}. 
Consequently, these datasets fall short in capturing the complex social and behavioral contexts encountered in real-world UserSD tasks.
Thus, there is a pressing need for a new, high-quality benchmark that captures both linguistic and social dimensions at the user level.

To address these challenges, we introduce \textbf{TwiUSD}, the first manually annotated user-level stance detection dataset with explicit social structure, which is the largest dataset comprising 16,211 users and 47,757 tweets.
Unlike previous datasets, TwiUSD explicitly captures social structure by incorporating follow relationships from the Twitter network. Importantly, stance labels are determined by considering both users' own tweet content and the stances of their followees, enabling TwiUSD to reflect both linguistic and structural indicators. This makes TwiUSD a more realistic and challenging benchmark for evaluating user-level stance detection models.

A key challenge in UserSD is that a user’s stance often cannot be determined from their own tweets alone; leveraging the social context—such as the stances of followees—is essential, yet introduces substantial noise and irrelevant information. 
Existing approaches typically aggregate information from all neighbors, which can amplify noise and reduce robustness.
To address this, we propose \textbf{MRFG}, a \textit{Multi-scale Relevance Filtering for Graph-aware Stance Detection} framework that selectively identifies stance-relevant signals from a noisy social context. MRFG consists of two main components: the \textit{Multi-scale Relevance Filter} module, which uses a large language model to assess the semantic relevance of followee tweets and filter out noisy content. It also ranks feature dimensions based on structural informativeness. The \textit{Graph-Sensitive Inference} module splits features into structure-sensitive and structure-neutral groups processed via GNN and MLP, respectively. This dual-path design enables MRFG to jointly model semantic content and relational structure for accurate user-level stance prediction.



Our main contributions are as follows:

(1) We present TwiUSD, the first manually annotated user-level stance detection dataset to explicitly incorporate social structure. With 16,211 users, it is also the largest of its kind, setting a new and realistic benchmark for the field.
    
(2) We propose {MRFG}, a relevance-aware and structure-informed framework that effectively filters noisy context and jointly leverages semantic and social information for stance prediction.
    
(3) Comprehensive experiments on TwiUSD demonstrate that MRFG significantly outperforms strong baselines, validating the effectiveness of context selection and structure-aware modeling. We will release TwiUSD and code upon acceptance to facilitate further research and reproducibility.

\section{Related Work}

\textbf{Stance Detection Datasets.}
Early work on stance detection primarily focuses on the tweet level. SemEval 2016 (SEM16)~\cite{MohammadKSZC16} is a widely used benchmark containing labeled tweets toward specific targets, followed by domain-specific datasets such as P-Stance~\cite{LiSSNIC21} and COVID-19 stance~\cite{glandt2021stance}. Some recent efforts also explore conversational stance detection, modeling stance across post-reply threads~\cite{villa2020stance,li2022improved, li2023contextual}.
In contrast, user-level stance detection aims to predict a user’s overall stance by aggregating multiple signals. To reduce annotation costs, many existing UserSD datasets rely on unsupervised or heuristic-based methods, such as retweet behavior~\cite{DBLP:conf/icwsm/DarwishSAN20, samih-darwish-2021-topical}, shared hashtags~\cite{DBLP:conf/icwsm/ZhangZP024},  aggregated tweet timelines~\cite{10.1007/978-3-031-45275-8_7, 10.1007/978-3-031-17091-1_43} or following~\cite{ZHU2020102031}. However, these approaches often suffer from noisy labels and ignore structural signals such as user relationships or profile context. Developing high-quality, manually annotated UserSD datasets that incorporate social structure is essential for advancing research in realistic stance detection scenarios.


\textbf{Stance Detection Approaches.}
Early stance detection methods focused on capturing target-specific cues through attention-based architectures~\cite{dey2018topical} and graph-based approaches~\cite{LiPLSLWYH22} that model relationships between text and targets. With the rise of pre-trained language models (PLMs) such as BERT~\cite{devlin-etal-2019-bert}, fine-tuning and prompt-based techniques~\cite{shin2020autoprompt, liang2022jointcl, KEPrompt} have become mainstream.
More recently, large language models (LLMs) have enabled context-aware prompting~\cite{zhang2023investigating} and human-in-the-loop learning~\cite{cai2023human}, leading to advances in stance detection. Work such as COLA~\cite{lan2024stance} and GraphICL~\cite{sun-etal-2025-graphicl} incorporates LLMs into multi-perspective and graph-structured scenarios. 

For user-level stance detection, existing work often aggregates tweet-level predictions~\cite{samih-darwish-2021-topical}, applies clustering~\cite{DBLP:conf/icwsm/DarwishSAN20}, or relies on LLM-driven tweet scoring~\cite{10.1007/978-3-031-45275-8_7}. Others use graph-based models~\cite{DBLP:conf/icwsm/ZhangZP024} to encode user-tweet interactions. 
However, many methods rely on heuristics, overlook relevance filtering, or underuse structural features. This underscores the need for refined models and datasets that better integrate textual and social signals for robust UserSD.

\section{TwiUSD Dataset}
\subsection{Data Collection}

To ensure the credibility, scale, and structural richness of user-level stance data, we built upon the TwiBot-22 dataset~\cite{feng2022twibot}, a large-scale Twitter benchmark containing nearly 1 million users with tweet histories, profile descriptions, and social connections—making it ideal for network-aware stance research (see Appendix~\ref{TwiBot-22:appendix}).

We identified the 2020 U.S. presidential election as the most prominent theme in TwiBot-22 based on hashtag frequencies and tweet volumes, with emotionally polarized discussions around \textit{Joe Biden} and \textit{Donald Trump}. Thus, these two candidates were selected as primary stance targets.
To ensure annotation quality and topic relevance, we filtered out all bot accounts, retaining only authentic, human users. We then extracted tweets that (1) explicitly mention the target (\textit{Biden} or \textit{Trump}) and (2) include election-related hashtags. Hashtag selection followed prior work~\cite{kawintiranon-singh-2021-knowledge, liang-etal-2024-multi}, focusing on politically salient yet ideologically neutral terms to minimize annotation bias. The complete hashtag list and filtering strategies are detailed in Appendix~\ref{keyword:appendix}.
The resulting data comprise two target-specific subsets (see Table~\ref{tab:stage_count}, Stage-1), providing a high-quality foundation for user-level stance annotation and further analysis.

 \begin{table}[htbp]
 
 \centering
 \fontsize{9pt}{12pt}\selectfont
  \begin{tabular}{cccc}
    \hline
     \textbf{Stage}&\textbf{Target} &  \textbf{User}&\textbf{Tweet}\\
    \hline
  \multirow{2}{*}{\textbf{Stage-1}}&\textbf{Biden} &  26,584&58,783\\
  &\textbf{Trump}   & 24,967&68,338\\
 \hline
 \multirow{2}{*}{\textbf{Stage-2}}& \textbf{Biden} & 8,348&19,084\\
 & \textbf{Trump}   & 10,837&31,253\\
   \hline
  \end{tabular}
 \caption{\label{tab:stage_count} Entries after hashtag filter.}
\end{table}

\subsection{Data Preprocessing}

To enable effective stance reasoning, we designed a multi-step preprocessing pipeline to construct a user graph with both strong structural connectivity and semantic diversity.

(1) We first built the initial user graph by extracting all \textit{follow} relationships among filtered users, distinguishing between followees (being followed) and followers (following others). This captures the most meaningful and persistent ties for modeling social influence~\cite{ZHU2020102031}.

(2) To densify the network and enhance structural cues, we further added users who follow the identified followees, resulting in a more interconnected and realistic social graph and alleviating the sparsity common in prior datasets.

(3) To enrich semantic coverage, we included a small number of highly active users without follow links but with substantial tweet volumes, ensuring greater textual diversity and supporting model generalization.

The final user relationship statistics after preprocessing are summarized in Table~\ref{tab:stage_count} Stage-2. Specifically, the number of followees identified for the \textit{Biden} and \textit{Trump} targets was 1,010 and 1,113, respectively. In total, the dataset includes 20,468 follow links among users, forming a structurally rich graph for stance modeling.

\subsection{Data Annotation and Quality Assurance}

We adopted a three-class annotation scheme for user stance labels: \textit{Favor}, \textit{Against}, and \textit{None}, indicating support, opposition, or no clear attitude toward the target, respectively. 
Unlike prior datasets that treat all users uniformly, our annotation strategy explicitly leverages the underlying follow network structure. 
Users are divided into followees, followers, and isolated individuals, and annotation protocols are designed accordingly.

First, followees and isolated users are labeled based on manual inspection of their tweets and profile descriptions, reflecting their direct communication and self-presentation. Then, for follower users, stance labels are inferred by jointly considering their own content and the stances of their followees, mirroring realistic stance propagation patterns in social networks.

Rigorous quality assurance procedures are implemented throughout:
(1) {Qualified annotators}: Eight annotators with verified domain knowledge participated, each required to pass a trial annotation phase reviewed by two senior adjudicators. (2) {Blinded double annotation and multi-stage adjudication}: Each instance was independently labeled by two annotators. Disagreements (16\% of instances) triggered a second round of annotation, where a third-party adjudicator rendered the final decision. All annotation was performed independently and blindly, with no direct communication among annotators. 

To quantify reliability, we computed Cohen’s kappa statistic~\cite{kappa} and overall inter-annotator agreement, using the ``Favor'' and ``Against'' classes as in~\cite{LiSSNIC21}. 
The resulting kappa scores for the \textit{Biden} and \textit{Trump} targets were 0.90 and 0.91, respectively, indicating near-expert agreement and high annotation quality. Compared to prior datasets relying on distant supervision or heuristics, our annotation pipeline ensures reliable and realistic stance labels, laying a solid foundation for downstream modeling.

\subsection{Data Analysis}

\begin{figure}[htbp]
\centering
\includegraphics[width=0.85\linewidth]{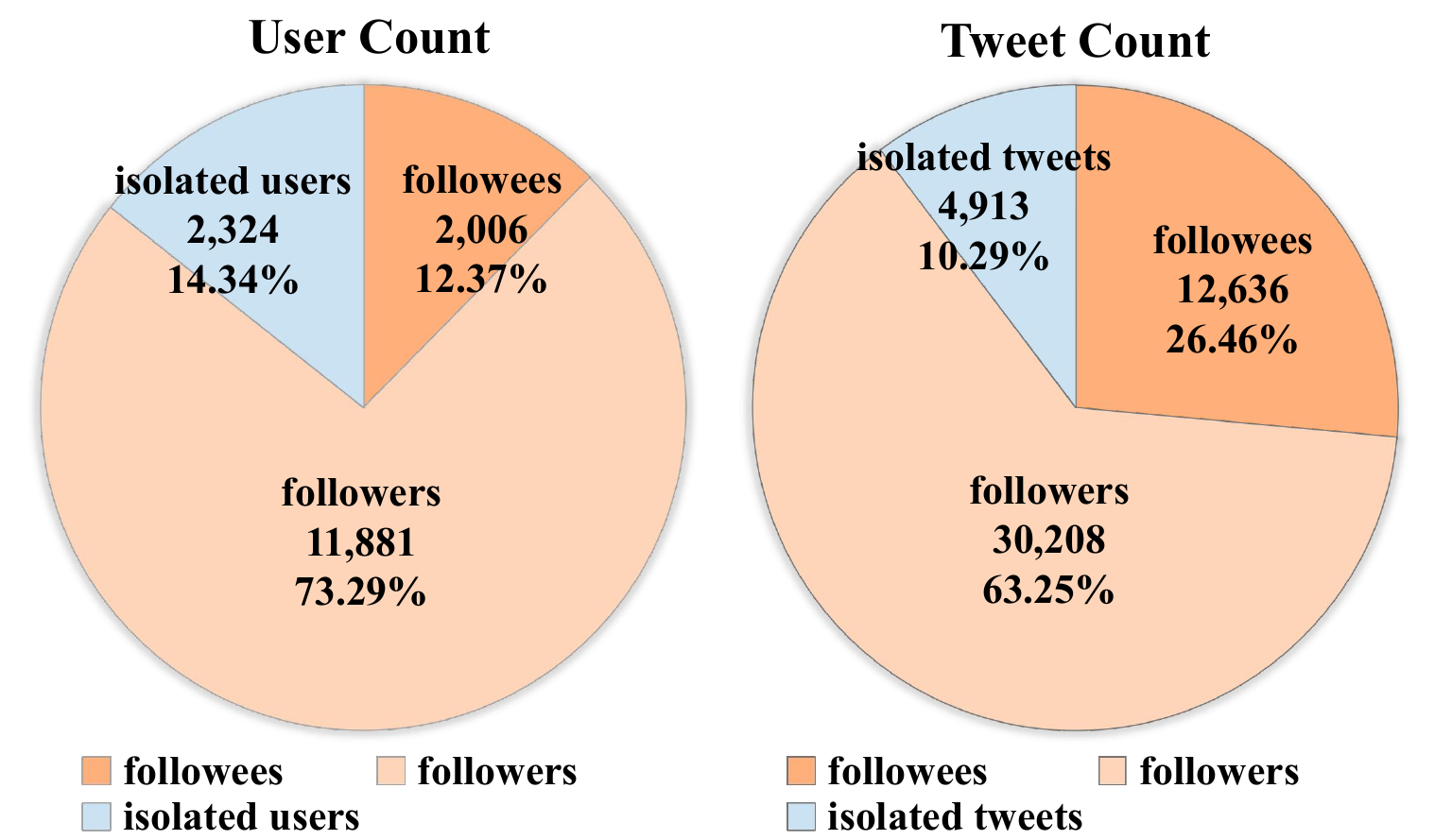}
\caption{User and tweet distributions by category in the TwiUSD dataset.} 
\label{fig:user-tweet}
\end{figure}

The TwiUSD dataset comprises 16,211 users and 47,757 tweets. As shown in Figure~\ref{fig:user-tweet}, the majority of users are followers (73.29\%), with followees and isolated users representing 12.37\% and 14.34\%, respectively. Despite their smaller numbers, followees contribute over a quarter of all tweets (26.46\%), indicating significantly higher activity levels and central roles in information dissemination. Table~\ref{tab:TwiUSD} details the stance label distribution. This multi-faceted user composition and activity heterogeneity provide a realistic testbed for stance modeling, and surpass the diversity of most existing user-level stance datasets.

To assess the reliability of hashtag-based unsupervised stance labeling, we applied two representative methods that infer user stance from stance-indicative hashtags. Our analysis reveals that these approaches suffer from substantial misclassification rates, with 56.16\% of users incorrectly labeled with the opposite stance. This finding underscores the limitations and potential biases of hashtag-based supervision, and highlights the necessity of manual, context-sensitive annotation for robust user-level stance detection. Detailed comparison results are provided in Appendix~\ref{Comparison dataset:appendix}.
For supervised experiments, we partitioned the dataset for each target into training, validation, and test sets with a 70/15/15 split, ensuring robust evaluation and future reproducibility.

\begin{table}[htbp]
\fontsize{9pt}{12pt}\selectfont
\begin{center}
    \resizebox{\linewidth}{!}{
\begin{tabular}{cccccccc}
\hline
 \multirow{2}{*}{\textbf{Target}}& \multicolumn{7}{c}{\textbf{Samples and Proportion of Labels}} \\
 \cline{2-8} 
 & Against& \%& Favor& \%& None&\%  &Total
  \\
  \hline
 \textbf{Biden}& 1,360& 19.76& 4,110& 59.70
& 1,414& 20.54
& 6,884 \\
 \textbf{Trump}& 6,089& 65.28& 1,546& 16.58
& 1,692& 18.14
&9,327 \\
 \hline
 \textbf{Total}& 7,449& 45.95& 3,106& 34.89& 5,656& 19.16&16,211 \\
\hline
 \end{tabular}
 }
\caption{\label{tab:TwiUSD}Label distribution of the TwiUSD dataset.}
 \end{center}
\end{table}

\section{Methods}

\begin{figure*}[htbp]
\centering
\includegraphics[width=0.9\linewidth]{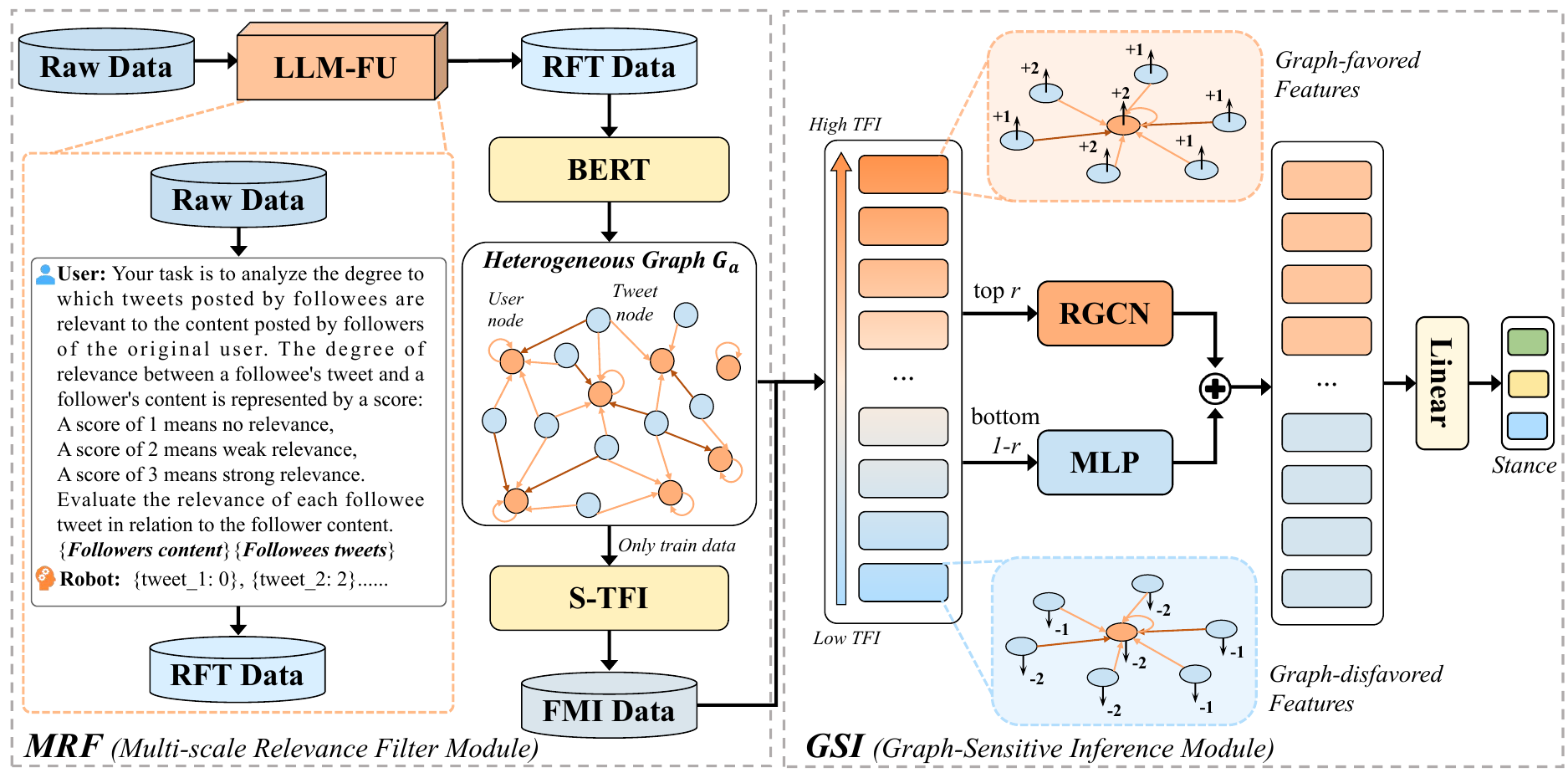}
\caption{The architecture of our MRFG framework.} 
\label{fig:model}
\end{figure*}

\subsection{Task Definition}
Given a user $u$, our goal is to predict the user's stance $y_u \in \{ \text{favor}, \text{against}, \text{none} \}$ towards a specific target (e.g., \textit{Trump} or \textit{Biden}). Each user is associated with two types of content: their own tweet collection $T_u$ and the tweet set of users they follow (i.e., followees) $T_{F(u)}$. In addition, we use the user's profile description $d_u$ and structural interactions (following) to enhance representation learning. Formally, the stance detection task is defined as:
\begin{equation}
f(u) = \arg\max_{y \in \mathcal{Y}} P(y \mid T_u, T_{F(u)}, d_u, G_a)
\end{equation}
where $G_a$ denotes the social interaction graph, and $\mathcal{Y}$ is the label space.

\subsection{Framework Overview}

Our proposed MRFG framework jointly leverages social structure and textual content to robustly infer user stance while filtering out noisy context. As illustrated in Figure~\ref{fig:model}, MRFG consists of two main modules:
MRF (Multi-scale Relevance Filter) extracts the most relevant social and textual signals by: (1) using an LLM-based filter to select followee tweets highly relevant to the target user, and (2) ranking feature dimensions using the TFI metric~\cite{zheng2025let}, which quantifies how well each feature aligns with the graph structure to support stance reasoning.
GSI (Graph-Sensitive Inference) processes the selected features in two paths: a relational graph convolutional network (RGCN)~\cite{schlichtkrull2017modelingrelationaldatagraph} for structure-sensitive features, and a simple multi-layer perceptron (MLP) for content-sensitive features. The fused outputs are used for final stance prediction.
This hybrid design allows MRFG to adaptively balance content and structural cues, outperforming single-path or homogeneous modeling approaches.

\subsection{MRF Module}

A user's stance often cannot be inferred accurately from their profile description and personal tweets alone, making it necessary to incorporate social context. However, directly utilizing all followee tweets introduces considerable noise, as not all content is relevant to the user's stance. To address this, we propose MRF, which identifies informative signals by filtering at both the tweet level and the feature dimension level. Inspired by prior work on structure-aware feature selection~\cite{zheng2025let}, MRF further assesses which features are most suitable for graph-based learning, thereby enabling more effective stance reasoning.
MRF operates in two main steps:

\textbf{(1) Relevance Filtering via LLM-FU.}
We first employ a LLM-based Filtering Unit (LLM-FU) to evaluate the semantic relevance between each followee tweet $T_{F(u)}$ and the target user's content. Prompt templates are designed to guide the LLM in assigning each tweet a relevance score in \{1, 2, 3\}, representing \textit{none}, \textit{weak}, and \textit{strong} relevance, respectively. Tweets scored as 1 (none) are discarded, while those with scores $\geq 2$ are retained as the Relevance-Filtered Tweets (RFT) Data. Prompt templates and examples are shown in Appendix~\ref{prompt:appendix}.

\textbf{(2) Feature Ranking via S-TFI.}
Given the filtered RFT Data, we proceed to extract and rank feature dimensions for each user $u$. We encode the user's profile description $d_{u}$ and their tweet collection $T_{u} = \{ T_{u}^1, \ldots, T_{u}^i \}$, where $i$ is the number of tweets posted by user $u$. These elements are concatenated with special separator tokens to form the input sequence.

Given the filtered RFT data, we proceed to extract and rank features. For each user $u$, we encode their profile description $d_{u}$ and their tweet collection $T_{u} = { T_{u}^1, \ldots, T_{u}^n }$. These elements are concatenated into a single input sequence using a special separator token, resulting in: {[CLS]} $d_{u}$ {[SEP]} $T_{u}^1$ {[SEP]} $\ldots$ $T_{u}^n$ {[SEP]}. This sequence is fed into a BERT encoder, and mean pooling is applied over the token embeddings to obtain the user representation $E_{u}$. For each followee tweet $t$ in $T_{F(u)}$, we similarly use BERT to generate its corresponding embedding $E_{t}$.
Next, we construct a directed heterogeneous graph $G_a$, where tweet nodes point to user nodes. Each user node is also assigned a self-loop to preserve self-referential information. 

Using this graph structure, we apply the S-TFI (Structural Topological Feature Informativeness) framework to assess the importance of each individual feature dimension for graph-based learning. S-TFI measures how well a feature dimension aligns with the graph topology in contributing to node label discrimination.
Given user node features $X \in \mathbb{R}^{n \times d}$ and node labels $Y$, we compute the informativeness of each feature dimension $m$ as:
\begin{equation}
\text{TFI}_m = I(Y; \tilde{X}_{:,m}) \quad \text{with } \tilde{X} = \hat{A} X
\end{equation}
where $\hat{A}$ is the normalized adjacency matrix of the graph, and $\tilde{X}$ represents the smoothed features after propagation through the graph. The mutual information $I(Y; \tilde{X}_{:,m})$ quantifies how informative the $m$-th feature is for predicting node labels, considering the structural context.

Based on the TFI scores, we sort all feature dimensions in descending order of informativeness. The resulting ranked feature set is referred to as the feature Mutual Information(FMI Data), which is used in the stance inference module for guided feature selection. To ensure fairness and prevent data leakage, the FMI Data is computed exclusively using node features from the training set, without incorporating any information from validation or test data.

\subsection{GSI Module}

The GSI module performs user-level stance detection based on the FMI Data generated by the MRF module. Rather than treating BERT features as static, our framework trains the entire pipeline end-to-end, allowing both the textual encoder and graph-based reasoning components to co-adapt during optimization.

We first split the user features into two disjoint subsets based on their TFI scores: the top Ratio($r$) features are selected as \textit{graph-favored features} $X_G$, while the remaining are treated as \textit{graph-disfavored features} $X_{\neg G}$. These two subsets are processed via separate encoders.

\textbf{(1) Graph-Favored Encoding via RGCN.} 
We use a RGCN~\ to process high-value features from user nodes over the heterogeneous user-tweet graph $G = (V, E, \mathcal{R})$, where $\mathcal{R}$ denotes relation types (e.g., tweet-to-user, self-loops). 

Specifically, we select the top-$r$ TFI-ranked feature dimensions from user node embeddings as \textit{graph-favored features}, denoted by $X_G$. These features are used as the initial input to the RGCN, capturing structural dependencies among users and their related content.

Formally, given initial feature representations $h^{(0)} = X_G$, the hidden representation at layer $l+1$ for user node $i$ is computed as:
\begin{equation}
\small
h_i^{(l+1)} = \sigma\left( \sum_{\zeta \in \mathcal{R}} \sum_{j \in \mathcal{N}_i^\zeta} \frac{1}{c_{i, \zeta}} W_\zeta^{(l)} h_j^{(l)} + W_0^{(l)} h_i^{(l)} \right)
\end{equation}
where $\mathcal{N}_i^\zeta$ denotes the set of neighbors of user node $i$ under relation $\zeta$, $c_{i,\zeta} = |\mathcal{N}_i^\zeta|$ is a normalization constant, and $W_\zeta^{(l)}$, $W_0^{(l)}$ are trainable parameters. $\sigma(\cdot)$ is a non-linear activation function. We use a 2-layer RGCN to obtain the final structure-aware representation $Z_G$ for each user node.

\textbf{(2) Graph-Disfavored Encoding via MLP.}
The graph-disfavored features $X_{\neg G}$ are processed using a MLP without any graph-based propagation. We adopt a 2-layer MLP with ReLU activation functions to transform these features and capture their semantic representations independently of the graph structure. The resulting output is denoted as $Z_{\neg G}$.

\textbf{(3) Feature Fusion and Stance Prediction.}
The outputs of the two encoding paths are concatenated and passed to a final linear classifier:
\begin{equation}
Z = [Z_G \, || \, Z_{\neg G}], \quad \hat{y} = \text{Linear}(Z)
\end{equation}

We use the standard cross-entropy loss for stance classification.

\section{Experimental Setup}

\textbf{Evaluation Metrics.}
We use $F_1$ score and accuracy ($Acc$) to evaluate model performance, following~\citet{LiSSNIC21} and~\citet{10.1145/3003433}. Specifically, we report $F_{\text{favor}}$ and $F_{\text{against}}$ for the ``\textit{Favor}'' and ``\textit{Against}'' classes, their average $F_{\text{avg}}$, and overall accuracy.

\begin{table*}[t]
\small
\centering
\fontsize{8pt}{10pt}\selectfont
  \resizebox{0.8\linewidth}{!}{
  \begin{tabular}{lccclclrrrlr}
    \hline
     \multirow{2}{*}{\textsc{METHOD}}&      \multicolumn{5}{c}{Biden}&&\multicolumn{5}{c}{Trump}\\
    \cline{2-6} \cline{8-12}& $F_{\text{favor}}$&$F_{\text{against}}$&$F_{\text{avg}}$ && $Acc$&&$F_{\text{favor}}$&$F_{\text{against}}$&$F_{\text{avg}}$ &&$Acc$\\
    \hline
     UUSDT&      31.63&38.96&35.29 && 36.14&&12.15&64.50&38.32 &&48.53\\
  Tweets2Stance& 12.18& 47.37& 29.77 && 32.78&& 39.76& 88.43& 64.09 &&79.72\\
   \hdashline
  TAN
& 66.45 & 12.54 & 39.49  &&  50.24 && 47.63 & 84.19 & 65.91  &&74.18 \\
  CrossNet& 68.55 & 35.59 &  52.07  && 55.39 && 33.90 & 79.97 &  53.60  &&66.63 \\
  \hdashline
  BERT
& 82.41 & 69.49 & 75.95  && 75.92 
&& 62.95 & 87.52 & 75.23  &&79.57 
\\
 RoBERTa& 83.72 & 74.69 & 79.21  && 77.72 && 64.66 & 88.03 & 76.34  &&79.61 \\
  BERT-GCN& 65.94 & 41.91 & 53.92  && 52.39 && 30.67 & 79.42 & 55.05  &&63.40 \\
  TPDG& 80.48 & 67.90 & 74.19  && 74.17 && 58.70 & 86.86 & 72.78  &&78.35\\
  JointCL
& 81.12 & 70.83 & 75.97  && 74.89&& 56.21 & 86.40 & 71.31  &&77.68\\

  KEPrompt& 80.31 & 65.61 & 72.96  && 72.55 && 57.08 & 87.29 & 72.19  &&78.14 \\
   \hdashline
  LLaMA2-70B& 77.52& 65.15& 71.34 && 65.21
&& 42.95& 77.11& 60.03 &&59.19
\\
  LLaMA3-70B& 66.30 & 56.58 &  61.44  && 52.36 && 58.77 & 73.08 &  65.92  &&59.34 \\
  GPT-3.5& 72.36& 58.85& 65.61 &&  60.37 && 55.63 & 84.64 & 70.13  &&72.18 \\

  GPT-4& 61.16& 65.65& 63.41 &&57.27&&59.17&70.52& 64.85 &&61.25 \\
  COLA &75.48&60.78&68.13 &&62.94&&55.50 &84.16 &69.83  &&72.10 \\
 GraphICL& 75.53 & 55.99 & 65.76  && 63.18 & & 46.38 & 89.12 & 67.75  &&81.27 \\
 
   \hdashline
 
 \textbf{\cellcolor[gray]{0.92} MRFG }&\cellcolor[gray]{0.92}\textbf{88.80} &\cellcolor[gray]{0.92}\textbf{79.57} &\cellcolor[gray]{0.92}\textbf{84.19}  &\cellcolor[gray]{0.92}
&\cellcolor[gray]{0.92}\textbf{83.04} &\cellcolor[gray]{0.92}&\cellcolor[gray]{0.92}\textbf{69.74} &\cellcolor[gray]{0.92}\textbf{92.80} &\cellcolor[gray]{0.92}\textbf{81.27}  &\cellcolor[gray]{0.92}
&\cellcolor[gray]{0.92}\textbf{87.47} \\
 
  \cellcolor[gray]{0.92} {\textit{w/o}} \ LLM-FU&\cellcolor[gray]{0.92}86.58 &\cellcolor[gray]{0.92}76.08 &\cellcolor[gray]{0.92}81.33  &\cellcolor[gray]{0.92}
&\cellcolor[gray]{0.92}80.78 &\cellcolor[gray]{0.92}&\cellcolor[gray]{0.92}63.67 &\cellcolor[gray]{0.92}92.41 &\cellcolor[gray]{0.92}78.04  &\cellcolor[gray]{0.92}
&\cellcolor[gray]{0.92}85.53 \\
  
   {\cellcolor[gray]{0.92} \textit{w/o}} \ $S$-$TFI_R$&\cellcolor[gray]{0.92}88.48&\cellcolor[gray]{0.92}79.29 &\cellcolor[gray]{0.92}83.88  &\cellcolor[gray]{0.92}
&\cellcolor[gray]{0.92}82.96 &\cellcolor[gray]{0.92}&\cellcolor[gray]{0.92}67.57 &\cellcolor[gray]{0.92}92.80 &\cellcolor[gray]{0.92}81.27  &\cellcolor[gray]{0.92}
&\cellcolor[gray]{0.92}86.20 \\
   
{\cellcolor[gray]{0.92} \textit{w/o} \ $S$-$TFI_m$}&\cellcolor[gray]{0.92}86.39 &\cellcolor[gray]{0.92}77.18 &\cellcolor[gray]{0.92}81.78  &\cellcolor[gray]{0.92}&\cellcolor[gray]{0.92}81.24 &\cellcolor[gray]{0.92}&\cellcolor[gray]{0.92}66.67 &\cellcolor[gray]{0.92}91.72 &\cellcolor[gray]{0.92}79.20  &\cellcolor[gray]{0.92}&\cellcolor[gray]{0.92}87.03 \\
\hline
  \end{tabular}
  }
\caption{ \label{tab:in-target}In-target experimental results (\%) on the TwiUSD dataset. The best performance in each group is highlighted in bold. Results for MRFG are reported with feature selection ratio $r = 0.3$.}

\end{table*}

\textbf{Baseline Methods.}
We compare our method with representative baselines from four categories: UserSD methods, supervised neural models, fine-tuned pre-trained models, and LLM-based prompting methods.
\textit{UserSD methods:}
\textbf{UUSDT}\cite{DBLP:conf/icwsm/DarwishSAN20},
\textbf{Tweets2Stance}\cite{10.1007/978-3-031-45275-8_7};
\textit{Supervised Neural Models:}
\textbf{TAN}\cite{ijcai2017p557},
\textbf{CrossNet}\cite{xu2018cross};
\textit{Fine-Tuned PLMs:}
\textbf{BERT}\cite{devlin-etal-2019-bert},
\textbf{RoBERTa}\cite{DBLP:journals/corr/abs-1907-11692},
\textbf{BERT-GCN}\cite{liu2021enhancing},
\textbf{TPDG}\cite{LiangF00DHX21},
\textbf{JoinCL}\cite{liang2022jointcl},
\textbf{KEPrompt}\cite{KEPrompt};
\textit{LLM-Based Methods:}
\textbf{LLaMA 2/3},
\textbf{ChatGPT (3.5/4)},
\textbf{COLA}\cite{lan2024stance},
\textbf{GraphICL}\cite{sun-etal-2025-graphicl}.
For a brief description of each baseline model, please refer to Appendix~\ref{baseline:appendix}.

\section{Experimental Results}

In this section, we perform comprehensive experiments on our TwiUSD dataset. The reported results are averaged over three runs with different random seeds to ensure robustness.

\subsection{In-Target Stance Detection}
We first report the experimental results on the TwiUSD dataset under the in-target setting, where both the training and testing instances share the same stance targets. The results are presented in Table~\ref{tab:in-target}. From these results, we draw the following key observations.

First, our proposed method MRFG achieves the best performance across all evaluation metrics on both targets, significantly outperforming all baselines. Specifically, MRFG achieves an $F_{avg}$ of 84.19\% and an accuracy of 83.04\% on the \textit{Biden} subset, and 81.27\% / 87.47\% on the \textit{Trump} subset. This confirms the effectiveness of our relevance-aware filtering and graph-informed dual-path inference.
Second, traditional supervised neural models such as TAN and CrossNet perform moderately, with clear performance gaps compared to structure-aware models. Although they incorporate target-specific attention, they lack structural reasoning crucial for capturing user-level stance.
Third, structure-augmented models such as TPDG and JoinCL do not consistently outperform content-based PLMs like BERT and RoBERTa. This suggests that simply adding structural or contrastive components may be insufficient. In contrast, MRFG’s explicit context filtering and feature-sensitive design lead to consistently better results.

Finally, we conduct paired two-tailed t-tests between MRFG and the strongest baselines (JoinCL and TPDG). The improvements of MRFG are statistically significant across all metrics ($p < 0.05$), confirming the robustness of our proposed design.

\subsection{Cross-Target Stance Detection}

\begin{table}[h]
\small
\centering
\fontsize{8pt}{10pt}\selectfont
  \resizebox{1\linewidth}{!}{
  \begin{tabular}{lccclc}
    \hline
     \multirow{2}{*}{METHOD}&      \multicolumn{5}{c}{Trump$\rightarrow$Biden}\\
    \cline{2-6} & $F_{\text{favor}}$&$F_{\text{against}}$&$F_{\text{avg}}$ && $Acc$\\
  \hline
  TAN
& 21.63 & \textbf{62.88} & 42.25  &&  \textbf{49.58}\\
  CrossNet
& 36.88 & 32.91 &  34.89  && 33.79\\
  BERT
& \textbf{55.24}& 39.75 & 47.49  && 49.13\\
 RoBERTa
& 52.26 & 41.70 & 46.98  && 45.39\\
  BERT-GCN& 31.20& 59.86& 45.43 && 45.98\\
  JointCL& 55.21 & 41.56 & \textbf{48.38} && 44.86\\
  KEPrompt& 52.70 & 42.10 & 47.40  && 45.69\\
 
 \textbf{MRFG}&48.87 &41.45 &45.16  &&43.24\\
\hline
 \multirow{2}{*}{METHOD}& \multicolumn{5}{c}{Biden$\rightarrow$Trump}\\

 \cline{2-6}& $F_{\text{favor}}$& $F_{\text{against}}$& $F_{\text{avg}}$ & & $Acc$

\\
\hline
 TAN
& 61.22 & 25.31 & 43.27  & & \textbf{47.16}
\\
 CrossNet
& 28.12 & 21.31 & 24.71  & & 25.27
\\
 BERT
& 29.95 & 37.16 & 33.56  & & 32.49
\\
 RoBERTa
& 31.93 & 50.83 & 41.38  & & 39.36
\\
 BERT-GCN& \textbf{61.75}& 36.27 & \textbf{49.03} & & 46.80
\\
 JointCL& 31.87 & 46.86 & 39.36  & & 37.37
\\
 KEPrompt& 31.32 & 46.07 & 38.70  & & 35.88
\\
 \textbf{MRFG}& 32.60 & \textbf{59.55}& 46.08  & & 45.62\\
 \hline
  \end{tabular}
  }
\caption{ \label{tab:Cross-target}Cross-target experimental results (\%) on the TwiUSD dataset. Results for MRFG are reported with feature selection ratio $r = 0.3$.}

\end{table}

We evaluate model generalization under the cross-target setting, where training and test data involve different stance targets. As shown in Table~\ref{tab:Cross-target}, all models experience a noticeable drop in performance, reflecting the challenge of transferring stance knowledge across targets.

While BERT and JoinCL achieve the best $F_{avg}$ or accuracy in at least one direction, our model MRFG remains competitive, showing stable performance and achieving the highest $F_{\text{against}}$ in the Biden$\rightarrow$Trump setting. However, its overall performance is limited by the use of TFI-based feature selection, which is optimized on the training target and may not transfer well. This reveals a trade-off between target-specific modeling and cross-target generalization.




\subsection{Ablation Study}

To evaluate the impact of each core component in MRFG, we conduct ablation experiments with three variants:
(1) \textbf{w/o} LLM-FU, which removes the relevance estimation module;
(2) \textbf{w/o} S-TFI\textsubscript{R}, which processes all features using RGCN; and
(3) \textbf{w/o} S-TFI\textsubscript{m}, which processes all features using MLP.
As shown in Table~\ref{tab:in-target}, removing the LLM-FU module leads to a consistent performance drop, confirming the value of filtering irrelevant followee tweets. Without LLM-FU, the model is exposed to noisy inputs, reducing both $F_{avg}$ and accuracy.
In addition, removing the S-TFI-based dual-path design and replacing it with a single encoder (either RGCN or MLP) also results in reduced performance. The \textit{w/o S-TFI\textsubscript{m}} setting performs worse than \textit{w/o S-TFI\textsubscript{R}}, indicating that graph-based reasoning is more beneficial for structure-sensitive features, while unfiltered processing weakens the model’s ability to exploit feature informativeness.
These results validate the effectiveness of both selective relevance filtering and TFI-guided feature separation.

\subsection{Why use LLM for relevance estimation?}

\begin{figure}[htbp]
\centering
\includegraphics[width=1\linewidth]{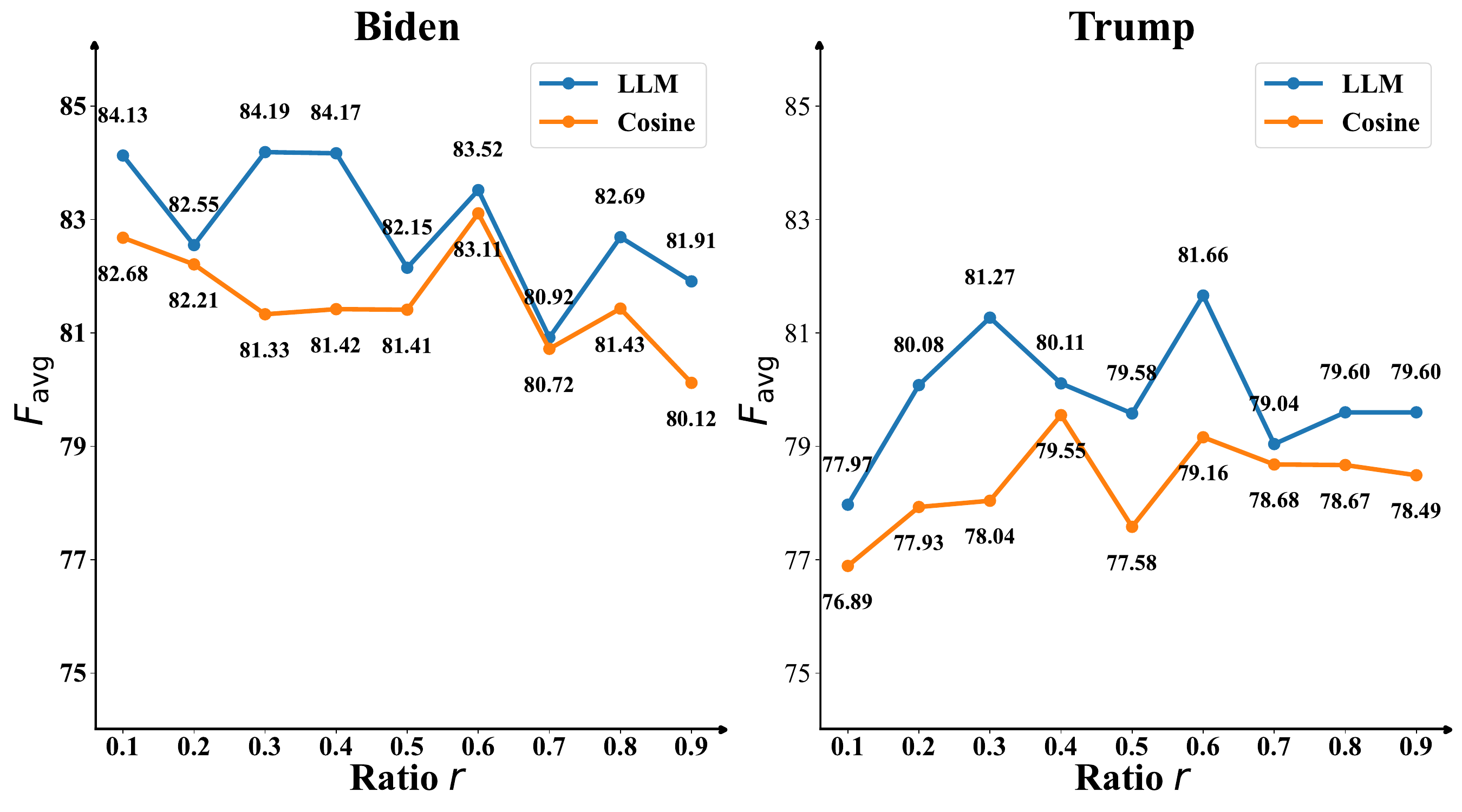}
\caption{Comparison of LLM-based vs. Cosine-based relevance estimation under different feature selection ratios $r$ on the TwiUSD dataset.} 
\label{fig:ratio-llm}
\end{figure}

To evaluate the effectiveness of our LLM-based Filtering Unit (LLM-FU), we compare it with a baseline that uses \textit{cosine similarity} between the BERT embedding of the target user's content and each followee tweet. In this baseline, tweets with cosine similarity in the range of $[0.7, 0.85)$ are treated as \textit{weakly relevant}, while those in $[0.85, 1]$ are considered \textit{strongly relevant}. Tweets below 0.7 are discarded as irrelevant.
As shown in Figure~\ref{fig:ratio-llm}, the LLM-based LLM-FU consistently outperforms the cosine-based approach across all feature selection ratios $r$ on both targets. This performance gap highlights the limitations of cosine similarity, which relies solely on shallow embedding similarity and cannot fully capture the nuanced semantic or pragmatic relevance between users and tweets.
In contrast, LLMs offer stronger context understanding and reasoning capabilities. They assess relevance not just based on textual similarity but also by interpreting stance-related intent and implicit references. This enables more accurate filtering of useful contextual tweets, ultimately leading to better stance prediction.

\subsection{Impact of Feature Selection Ratio \texorpdfstring{$r$}{r}}

We analyze the effect of the feature selection ratio $r$ used for splitting features between the GNN and MLP paths. Figure~\ref{fig:ratio-llm} presents $F_{avg}$ scores for both the LLM-based and cosine-based filtering approaches across varying $r$ values on the \textit{Biden} and \textit{Trump} subsets.

Across both targets and strategies, performance generally peaks when $r$ is in the range of $0.3$--$0.4$, indicating that a moderate allocation of structural features yields the most effective balance between content and relational signals. When $r$ is too small, the model lacks sufficient structural context; when $r$ is too large, overly relying on potentially noisy graph signals can degrade performance. Both filtering strategies follow a similar trend, though with slightly different sensitivities to extreme $r$ values.

\subsection{Error Analysis}
To understand the limitations of our model, we analyze misclassified users and identify two main sources of error. First, users with very limited tweet history, especially those with only one tweet, show a much lower prediction accuracy (approximately 79\%), which is far below the overall model performance. Second, users who follow many accounts with inconsistent or neutral content are also prone to errors, particularly when followees express conflicting stances. Detailed error statistics for both cases are provided in Appendix~\ref{appendix:error}.



\section{Conclusion}
We establish a new standard for user-level stance detection by introducing TwiUSD, the first large-scale, manually annotated benchmark with explicit social network structure. Our proposed MRFG framework, which combines LLM-based relevance filtering and structure-aware feature routing, achieves state-of-the-art performance across both in-target and cross-target settings. Comprehensive analyses demonstrate the value of selectively integrating social and linguistic signals for robust stance modeling. By releasing TwiUSD and our code, we aim to empower the community with a realistic testbed and effective tools for advancing research in opinion mining and social network analysis.




\clearpage

\section*{Limitations}
Our method is built entirely on textual content and structural relations derived from user tweets and followee interactions. The TwiUSD dataset only contains text-based information and does not include other modalities such as images, audio, or video, which may provide additional stance cues. Furthermore, our model does not incorporate any external data or knowledge sources for augmentation. It relies solely on the existing user-generated content without leveraging background knowledge, entity linking, or retrieval-enhanced mechanisms, which may limit its generalization in more open-domain or low-resource scenarios.

\section*{Ethics Statement}
Our dataset is constructed based on TwiBot-22, a publicly available Twitter dataset that has been released for academic research purposes. All data used in this study are obtained in accordance with the dataset’s license and comply with Twitter’s terms of service and privacy policies. We do not include any personally identifiable information beyond what is already publicly accessible in the original dataset.
We used the ChatGPT service from OpenAI for our writing. We followed their termss and policies.

\bibliography{custom}

\appendix

\begin{table*}[h]
\small
\centering
\fontsize{8pt}{10pt}\selectfont
    \begin{tabular}{c|c|c|c|c|c}
    \hline
    \textbf{Item}&\textbf{Value}&\textbf{Item}&\textbf{Value}&\textbf{Item}&\textbf{Value}\\
    \hline
    entity type&4&post&88,217,457&following&2,626,979\\
    relation type&14&pin&347,131&follower&1,116,655\\
    user&1,000,000&like&595,794&contain&1,998,788\\
    hashtag&5,146,289&mention&4,759,388&discuss&66,000,633\\
    list&21,870&retweet&1,580,643&bot&139,943\\
    tweet&88,217,457&quote&289,476&human&860,057\\
    user metadata&17&reply&1,114,980&entity&92,932,326\\
    hashtag metadata&2&own&21,870&relation&170,185,937\\
    list metadata&8&member&1,022,587&max degree&270,344\\
    tweet metadata&20&follow&493,556&verified user&95,398\\
    \hline
    \end{tabular}
    \caption{Statistics of TwiBot-22.}
    \label{tab:statistic}
\end{table*}

\begin{table*}[htbp]
\small
\fontsize{8pt}{10pt}\selectfont
\begin{center}
    \resizebox{\linewidth}{!}{
\begin{tabular}{ccccccccccccc}
\hline
 \multirow{2}{*}{\textbf{Method}}& \multirow{2}{*}{\textbf{Target}} & \multirow{2}{*}{\textbf{Predict Stance}} & \multicolumn{7}{c}{\textbf{Samples and Proportion of True Labels}} & & \multicolumn{2}{c}{\textbf{Error Predict}} \\
 \cline{4-10} \cline{12-13} 
 & & & Against& \%& Favor& \%& None&\%  &Total & & Count &\% 
   \\
  \hline
      \multirow{2}{*}{\textbf{UUSDT}}&\multirow{2}{*}{\textbf{Trump}} & \textbf{favor} &  128 & 79.50 & 7 & 4.35 & 26 & 16.15 & 161  & & 135 & 83.85\\
  & & \textbf{against} & 176 & 95.66 & 4 & 2.17 & 4 & 2.17 & 184  & & 8 & 4.34 \\
  \hdashline
  
\multirow{4}{*}{\textbf{DoubleH}}&\multirow{2}{*}{\textbf{Biden}} & \textbf{favor} &  18 & 6.41 & 24 & 8.54 & 239 & 85.05 & 281  & & 42 & 14.95 \\
 &  & \textbf{against} & 276 & 15.36 & 323 & 17.97 & 1198 & 66.67 & 1797  & & 1521 & 84.64 \\
    & \multirow{2}{*}{\textbf{Trump}} & \textbf{favor} & 2602 & 91.17 & 61 & 2.14 & 191 & 6.69 & 2854  & & 2663 & 93.31\\
   &    & \textbf{against} & 199 & 96.13 & 5 & 2.42 & 3 & 1.45 & 207  & & 8 & 3.86\\
   
\hline
 \end{tabular}
 }
\caption{\label{tab:Unsupervised}Analysis of hashtag-based stance prediction from unsupervised methods. We compare UUSDT~\cite{DBLP:conf/icwsm/DarwishSAN20} and DoubleH~\cite{DBLP:conf/icwsm/ZhangZP024}, which infer user stance from hashtag occurrences in tweets. The results show the distribution of true labels per predicted stance and the corresponding error rates.}
 \end{center}
\end{table*}
\begin{table*}[p]
    \centering
    \fontsize{8pt}{10pt}\selectfont
    \resizebox{\linewidth}{!}{
    \begin{tabular}{p{4cm}|p{12cm}}
        \hline
        \textbf{Role}&\textbf{Input}\\
        \hline
        \multirow{3}{*}{Instruction}&
        \multirow{3}{12cm}{Your task is to analyze the degree to which tweets posted by the original user are related to other posts. The degree of correlation between a post and a post is represented by a score, the score 1 means no association, the score 2 means weak association, and the score 3 means strong association.}\\
        \\  
        \\
        \hline
        \multirow{9}{*}{User's Tweets}&
        \multirow{9}{12cm}{1:"@MeidasTouch @realDonaldTrump Yes, there is nothing alike in the plots of daily new cases between the US and  other countries such as Germany.Just look at the shape, the peak (if there is any for the US), and most strikingly the scale (y-axis). \\
        2:"@drdavidsamadi @realDonaldTrump This is how the death cases look like when you reopen states too soon (see the figures). I can't imagine what will happen when you open all the schools. \\
        3:"@realDonaldTrump This argument is so shocking and anti-intellectual! The daily test positive ratio of many states such as TX is increasing. \\
        4:"@realDonaldTrump what".}\\
        \\ \\ \\ \\ \\ \\
        \\
        \\
        \hline
        \multirow{6}{*}{Followees's Tweets}&\multirow{6}{12cm}{348159742\_1:@fredyuksel @MsMaxwell6 @real-
        DonaldTrump i just got a FREEDOM BONER, \\
        3094891\_1: After close review of recent Tweets from the @realDonaldTrump account and \\ the context around them we have permanently su…, \\
        3094891\_2: Biden’s popular vote lead has grown to 6.3 million and 4\%:Biden
        80,301,58(51.1\%)Trump 73,978,678 (47.1\%), \\
        393705422\_1:@realDonaldTrump. }\\
        \\ \\ \\
        \\
        \\
        \hline
        \multirow{4}{*}{Output format }&\multirow{4}{12cm}{Use score to indicate the degree to which these tweets are related to the tweets posted by users, and are given in the order of tweets, only output the tweet number and corresponding score, the format example is "(tweet number:corresponding score)."}\\
        \\ \\ 
        \\
        \hline
        \multirow{1}{*}{Output data}&\multirow{1}{12cm}{(348159742\_1:1),
        (3094891\_1:2), 
        (3094891\_2:1), 
        (393705422\_1:2)}\\
        \hline
    \end{tabular}}
      \caption{\label{tab:Agent Example}Input structure of the large language model and the corresponding output examples.}
    \label{tab:hashtag}
\end{table*}

\begin{table*}[htbp]
\centering
\fontsize{8pt}{10pt}\selectfont
\resizebox{\linewidth}{!}{
\begin{tabular}{p{1cm}p{4cm}p{8cm}p{2cm}p{2cm}}
    \hline
    \textbf{Case} & \textbf{User's data} & \textbf{Followee's data} & \textbf{Predicted\_label} & \textbf{True\_label} \\
    \hline
    \multirow{1}{*}{1} & 
    tweet\_1: House Democrats watering down Biden's tax proposals: which to be fair, may be needed to get them passed… & 
    user1\_tweet1:10,000+ excess deaths a day caused by UK, Switzerland @EU\_Commission \& Germany blocking the \#TRIPSWaiver. \newline
    stance:[user1:'none'] & 
    none & 
    against \\
    \hline
    \multirow{1}{*}{2} & 
    tweet\_1: ``Putin may encircle Kiev with tanks, but he'll never gain the hearts and souls of the Iranian people.'' \#JoeBiden \#UkraineWar \#UkraineKrieg. & 
    user1\_tweet1: US Congressman asks President @JoeBiden to reject Pakistan's effort to install ‘jihadist’ envoy \newline
    Reported by: Sidhant Sibal (@sidh…) \newline
    stance:[user1:'against'] & 
    against & 
    none \\
    \hline
    \multirow{1}{*}{3} & 
    tweet\_1: BREAKING: \#SCOTUS will hear Biden v. Texas. This case will determine if the will of millions of voters who rejected Trump, \newline
    tweet\_2: \#DayWithoutImmigrants is trending at No.10 in US! \#Immigration \#WhiteHouse \#ImmigrationReformNow \#JoeBiden \#ValentinesDay… & 
    user1\_tweet1: Pres.-elect @JoeBiden nominated Merrick Garland as AG of @TheJusticeDept and other civil rights experts to DOJ. We look forward to working with the Biden/Harris DOJ to restore and enforce civil rights of America’s vulnerable farmworkers. \newline
    user1\_tweet2: President-elect @JoeBiden has nominated Tom Vilsack as the next @USDA Secretary. We are hopeful Vilsack will make the… \newline
    user2\_tweet1: If a so-called parole status is all that can get to President Biden’s desk, then Democrats need to stop playing games wi… \newline
    user2\_tweet2: HAPPENING NOW: TPS Families \& Community Faith Leaders in front of the White House for an action denouncing @joebiden, Dem… \newline
    ... \newline
    stance:[user1:'favor', user2:'against'] & 
    favor & 
    none \\
    \hline
    \multirow{1}{*}{4} & 
    tweet\_1: BREAKING NEWS: Trump’s Steve Bannon says that Americans should be supporting Russia instead of President Biden. RT IF… & 
    user1\_tweet1: If Democrats are worried about Biden looking weak, this is the kind of thing that actually makes him look weak because eve… \newline
    user1\_tweet2: Biden and Democrats getting the war with Russia they been begging for and trying to start since 2016. What’s wild is Ukrain… \newline
    user2\_tweet1: Both Republicans \& Democrats blast \#facialrecognition's threat to civil rights, so where's the ban?... \newline
    user3\_tweet1: I am profoundly honored to be the Principal Deputy Press Secretary for @JoeBiden. I am especially thrilled to work alongs… \newline
    user3\_tweet2: JUST IN: President-elect Joe Biden has officially been certified the winner of Georgia's 16 electoral votes following a statewide… \newline
    ... \newline
    stance:[user1:'against', user2:'none', user3:'favor'] & 
    favor & 
    against \\
    \hline
\end{tabular}}
\caption{\label{tab:case}Example of Error Case of Target ``Biden''.}
\end{table*}

\section{TwiBot-22 Information}
\label{TwiBot-22:appendix}

TwiBot-22 is a large-scale, high-quality annotated graph dataset, containing one million account information and tens of millions of tweets, as well as hundreds of millions of edges used to describe various types of edges. 
In addition, this dataset contains multiple social network relationships of users (such as following, like, forwarding, comments, etc.). This rich graph structure information can be used to analyze the relationships between users and the user's stance and tendency in social networks.
The specific dataset details are shown in Table \ref{tab:statistic}.
The dataset is large in scale and contains rich social network information, so it can be used to construct a social media account-level stance detection dataset and used for account-level stance detection tasks.

\section{TwiUSD Keywords}
\label{keyword:appendix}

To identify tweets relevant to the 2020 U.S. presidential election, we followed prior work~\cite{kawintiranon-singh-2021-knowledge, liang-etal-2024-multi} and selected a set of politically salient yet ideologically neutral hashtags. This approach avoids introducing bias during data collection while preserving political relevance.

We retained tweets that explicitly mentioned either ``\textit{Biden}'' or ``\textit{Trump}'' and contained at least one of the following hashtags:

\#Vote, \#Debates2020, \#USElection2020, \#Presiden-tialDebate2020, \#2020Election, \#votersuppression, \#GetOuttheVote, \#2020elections, \#Trump2020LandSlide, \#TrumpCrimeFamily, \#DonaldTrump, \#Republican, \#BidenForPresident, \#SleepyJoe, \#JoeBiden, \#Democrats

This keyword set was used to construct the initial tweet pool before user-level filtering and stance annotation.

\section{Comparison with unsupervised datasets}
\label{Comparison dataset:appendix}
Existing user-level stance detection methods often rely on unsupervised strategies, using stance-indicative hashtags to construct weakly labeled datasets.For example, 
UUSDT~\cite{DBLP:conf/icwsm/DarwishSAN20} labeled users who used
the hashtag \#MAGA as pro Trump and users who used any of the hashtags \#resist, \#resistance, \#impeachTrump, \#theResistance, or \#neverTrump as anti.

Besides,DoubleH~\cite{DBLP:conf/icwsm/ZhangZP024} used labeled users who used
the hashtag \#trumpvirus, \#bidenharris, \#biden2020, \#votebidenharris2020, \#resistant, \#votebluetoendthisnightmare, \#iamtulsi, \#trumpisanationaldisgrace, \#dumptrump, \#trumpmeltdown, \#voteblue as pro Biden, and \#gop, \#kag, \#hunterbiden, \#maga, \#trump2020, \#tcot, \#burisma, \#votered, \#republican, \#americafirst, \#sleepyjoe, \#mc2020, \#fourmoreyears, \#bidencrimefamily, \#bidenriots as pro Trump.

To assess the generalizability of these hashtag-based heuristics, we applied both UUSDT and DoubleH's labeling strategies to the TwiUSD dataset. The results, shown in Table~\ref{tab:Unsupervised}, indicate that these rules do not transfer well. In particular, users labeled by hashtag presence alone often hold the opposite stance, with error rates reaching up to 93.31\%. For example, when DoubleH predicted a user as ``favor Trump,'' 91.17\% of those users were actually ``Against.'' Similarly, UUSDT misclassified 83.85\% of users predicted as pro-Trump.

These findings suggest that stance-indicative hashtags, while useful in certain contexts, may be highly topic- and time-sensitive, and often fail to reflect the user's actual stance in broader datasets. This highlights the need for more robust, content- and structure-aware methods like ours for reliable user-level stance detection.

\section{LLM-FU Prompt}
\label{prompt:appendix}
We use an LLM-based filter to determine the correlation between tweets, which are used to elect followee tweets highly relevant to the target user.Specifically, we divide the tweet relevance into three categories: unrelated, weak correlation and strong correlation, which are expressed by scores 1, 2, and 3 respectively. For each tweet of the followees, a score is output to indicate the relevance to the target user's tweets. 
In our implementation, we use the GPT-4o model (via OpenAI API) as the underlying LLM for relevance assessment due to its strong contextual understanding and reasoning capabilities. The input format and representative output examples for the GPT-4o-based filtering process are shown in Table~\ref{tab:Agent Example}.

\section{Baseline Methods}
\label{baseline:appendix}

\paragraph{USD Methods.}

\textbf{UUSDT}~\cite{DBLP:conf/icwsm/DarwishSAN20} performs unsupervised stance detection by extracting user features and applying dimensionality reduction and clustering techniques.  
\textbf{Tweets2Stance}~\cite{10.1007/978-3-031-45275-8_7} leverages LLMs to identify tweet topics and estimate relevance, from which tweet-level stances are inferred and aggregated into user-level predictions.

\paragraph{Supervised Neural Models.}

\textbf{TAN}~\cite{ijcai2017p557} models the relationship between target and tweet using attention to learn target-specific stance representations.  
\textbf{CrossNet}~\cite{xu2018cross} performs cross-target stance detection by encoding tweets and targets via BiLSTM and applying aspect-level attention to extract stance-bearing components.

\paragraph{Fine-Tuned Pre-trained Models.}

\textbf{BERT}~\cite{devlin-etal-2019-bert} is fine-tuned directly on stance-labeled training data to serve as a strong contextualized baseline.  
\textbf{RoBERTa}~\cite{DBLP:journals/corr/abs-1907-11692} improves upon BERT by leveraging more training data, larger batch sizes, and optimized pre-training strategies.  
\textbf{BERT-GCN}~\cite{liu2021enhancing} enhances BERT with commonsense and structural knowledge using a graph convolutional layer for better generalization in low-resource settings.  
\textbf{TPDG}~\cite{LiangF00DHX21} dynamically disentangles target-dependent and target-independent components in stance expressions using structured modeling.  
\textbf{JoinCL}~\cite{liang2022jointcl} combines stance contrastive learning with target-aware graph contrastive learning to improve generalization to unseen targets.  
\textbf{KEPrompt}~\cite{KEPrompt} uses an automatic verbalizer to define label words and reformulates stance detection as a prompt-based classification task.

\paragraph{LLM-Based Prompting Methods.}

We implement LLM-based prompting baselines following the prompting strategy proposed by~\citet{gatto-etal-2023-chain}. We use \textbf{LLaMA 2-70B}~\cite{touvron2023llamaopenefficientfoundation}, \textbf{LLaMA 3-70B}, and \textbf{ChatGPT} (GPT-3.5 and GPT-4) to generate zero-shot or few-shot stance predictions via in-context learning.  
\textbf{COLA}~\cite{lan2024stance} introduces collaborative role-aware agents to model stance from multiple perspectives through LLM-based reasoning.  
\textbf{GraphICL}~\cite{sun-etal-2025-graphicl} incorporates graph-structured prompts into LLM input to enable relational stance reasoning over structured user-content graphs.

\section{Error Case}
\label{appendix:error}
To better understand the limitations of our model, we manually analyze several misclassified cases from the test set. We observe two major sources of error.

First, users with only a single tweet tend to be more difficult to classify. Specifically, we find that the prediction accuracy for users with only one tweet is 79\%, which is noticeably lower than the overall model accuracy of 85.26\%. These users often lack explicit stance cues, making both textual and structural signals insufficient for reliable inference (e.g., Table~\ref{tab:case} Case 1).

Second, users whose followees express conflicting or ambiguous stance signals are more difficult to classify. As shown in Case 3 and Case 4, the model sometimes over-relies on the dominant stance among followees, failing to reconcile contradictory signals—particularly when the user's own tweets are neutral or implicitly opinionated. In such scenarios, even LLM-based relevance filtering may preserve noisy or inconsistent inputs, leading to misclassification.

Table~\ref{tab:case} presents several illustrative error cases.
In Case 1, the user has a single tweet implying skepticism toward Biden’s tax proposal, noting that it may need revision to pass. Although this expresses indirect opposition, the limited content offers insufficient cues, leading the model to predict a neutral stance.
In Case 2, the user's tweet focuses on geopolitical issues unrelated to Biden, and no clear stance is expressed. However, all followees oppose Biden, which biases the model toward an incorrect ``against'' label.
In Cases 3 and 4, while the users' own tweets reveal their stance, the followees hold mixed and conflicting opinions. This inconsistency confuses the model's aggregation mechanism and leads to incorrect final predictions.

These findings suggest that future improvements should focus on disambiguating stance under sparse supervision and filtering conflicting social signals more robustly.


\end{document}